\PassOptionsToPackage{table}{xcolor}
\documentclass[sigconf]{acmart}

\usepackage{booktabs}
\usepackage{enumitem}
\usepackage{graphicx}
\usepackage{amsmath}
\usepackage{algorithm}
\usepackage{algpseudocode}
\usepackage{multirow}
\usepackage{placeins}
\usepackage{xcolor}

\newcommand{\runinhead}[1]{%
  \par\noindent\textbf{#1.}\enspace\ignorespaces}

\AtBeginDocument{%
  }

\copyrightyear{2027}
\acmYear{2027}
\setcopyright{cc}
\setcctype{by}
\acmConference[KDD '27]{Proceedings of the 33rd ACM SIGKDD Conference on Knowledge Discovery and Data Mining V.1}{August 01--05, 2027}{San Jose, CA, USA}
\acmBooktitle{Proceedings of the 33rd ACM SIGKDD Conference on Knowledge Discovery and Data Mining V.1 (KDD '27), August 01--05, 2027, San Jose, CA, USA}
\acmDOI{10.1145/3862868.3871496}
\acmISBN{979-8-4007-3134-1/2027/08}

\title{Multi-Decoder OneRec: Controllable Generative Retrieval for Multi-Objective Industrial Recommendation}

\newcommand{\cofirst}{\authornote{Equal contribution.}}
\newcommand{\cofirstmark}{\authornotemark[1]}
\newcommand{\corr}{\authornote{Corresponding author.}}
\newcommand{\corrmark}{\authornotemark[2]}

\newcommand{\ksaffil}{%
  \affiliation{%
    \institution{Kuaishou Technology}%
    \city{Beijing}%
    \country{China}%
  }%
}

\author{You Wang}\cofirstmark
\ksaffil
\email{wangyou05@kuaishou.com}

\author{Zhao Liu}\cofirstmark
\ksaffil
\email{liuzhao09@kuaishou.com}

\author{Guoping Tang}\cofirst
\ksaffil
\email{tangguoping@kuaishou.com}

\author{Yiqing Yang}
\ksaffil
\email{yangyiqing06@kuaishou.com}

\author{Shuo Su}
\ksaffil
\email{sushuo@kuaishou.com}

\author{Jing Liu}
\ksaffil
\email{liujing24@kuaishou.com}

\author{Naifu Zhou}
\ksaffil
\email{zhounaifu@kuaishou.com}

\author{Xiaoyou Zhou}
\ksaffil
\email{zhouxiaoyou@kuaishou.com}

\author{Wei Jiang}
\ksaffil
\email{jiangwei@kuaishou.com}

\author{Jian Liang}
\ksaffil
\email{liangjian03@kuaishou.com}

\author{Xiao Lv}\corr
\ksaffil
\email{lvxiao03@kuaishou.com}

\author{Ruiming Tang}\corrmark
\ksaffil
\email{tangruiming@kuaishou.com}

\author{Liyin Hong}
\ksaffil
\email{hongliyin@kuaishou.com}

\author{Wenwu Ou}
\ksaffil
\email{luocheng10@kuaishou.com}

\renewcommand{\shortauthors}{You Wang et al.}

\begin{document}

\begin{abstract}
Industrial recommender systems build candidate pools by assigning explicit
quotas to objective-specific retrieval routes. This design offers quota control
but increasingly fragments modeling, training, and serving as the route set
grows. Semantic-ID-based generative retrieval provides a unified alternative,
yet a single decoder entangles objective policies and limits candidate
complementarity. We propose \textbf{Multi-Decoder OneRec}, a controllable
framework that combines shared representations, isolated objective adaptation,
and coordinated decoding. All objectives share a user-context module and the
General Decoder, while each objective adds an isolated, parameter-efficient
LoRA expert. During training, exposure-sample next-token prediction (NTP)
updates the shared base, target-filtered NTP updates the event-based experts,
and Kullback--Leibler (KL)-regularized policy optimization updates the
Watch-time expert; gradient routing isolates these updates, and the General
Decoder supplies a stop-gradient reference. At inference, explicit route quotas
allocate the fixed budget and Multi-Decoder Constrained Beam Search reduces
cross-route overlap. We publicly release \textbf{Kwai26}, a large-scale
multi-objective benchmark with 1.31 billion raw item-level records, 31.85
million Item-ID entries, and 25.03 million items with valid Semantic IDs,
together with predefined splits and an evaluation protocol. Under the same
512-item retrieval budget, Multi-Decoder OneRec improves over the single-decoder
OneRec~\cite{deng2025onerec} baseline by 1.69\%--5.62\% across four Recall@512
metrics. In a production A/B test, it yields relative gains of 0.37\% in app
usage time per device, 0.19\% in Day-7 retained users, 0.19\% in devices with
at least one share, and 2.09\% in new-content Cold-Start. These results show
that generative retrieval can combine shared modeling with objective-specific
control and complementary candidate generation.
\end{abstract}


\maketitle

\section{Introduction}
\label{sec:introduction}

Candidate retrieval in large-scale recommender systems must allocate a fixed
candidate budget across multiple objectives~\cite{covington2016youtube,
huang2025multichannel}. Industrial systems combine numerous specialized
retrieval routes~\cite{xie2020ican,nie2022mic} for objectives such as watch
time, explicit interactions, and Cold-Start and assign each a
configurable budget share~\cite{huang2025multichannel,zhou2026capts}. This
design broadens coverage and directly controls candidate composition. However,
scaling the route set fragments model development, training, and serving while
increasing computational and maintenance overhead. These limitations motivate
a unified retriever that shares modeling capacity without sacrificing explicit
control over candidate composition.

\begin{figure}[t]
  \centering
  \includegraphics[width=\columnwidth,pagebox=cropbox]{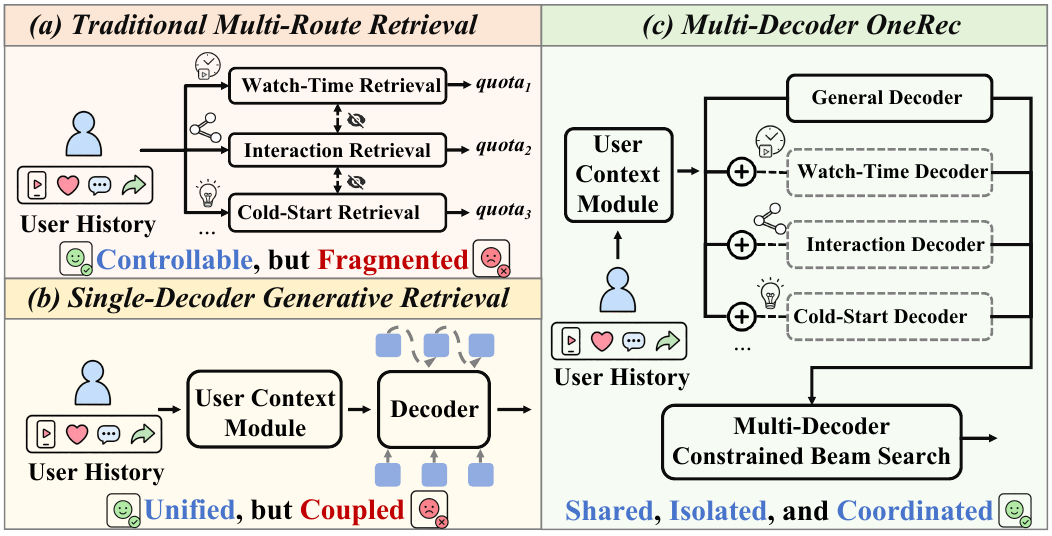}
  \caption{Comparison of multi-objective retrieval paradigms. (a) Traditional
  multi-route retrieval assigns quotas to independent routes, preserving quota
  control but fragmenting modeling, training, and serving. (b) Single-decoder
  generative retrieval shares modeling across objectives but couples their
  updates and may produce overlapping candidates. (c) Multi-Decoder OneRec
  shares the user-context module and General Decoder, isolates objective-specific
  experts, and coordinates quota-aware decoding with Multi-Decoder Constrained
  Beam Search (MD-CBS).}
  \Description{Three panels compare traditional multi-route retrieval,
  single-decoder generative retrieval, and Multi-Decoder OneRec. The proposed
  framework shares a user-context module and General Decoder, isolates
  objective-specific expert state, and coordinates candidates with
  Multi-Decoder Constrained Beam Search.}
  \label{fig:intro_overview}
\end{figure}

Generative retrieval provides the shared modeling needed for this goal by
representing each item as a discrete Semantic ID sequence and generating
candidates~\cite{rajput2023tiger,jin2024lmindexer,wang2024letter,
liu2026diffgrm}. Many unified generators serve multiple objectives with a
single decoder. Because all objectives use the same attention and feed-forward
layers, joint training couples their generation policies. An update for one
objective changes the transformations used by the others, which can cause
negative transfer when their supervision signals and preferred candidate
distributions diverge~\cite{mmoe,ple}. Existing methods inject objective
information through prompts, task tokens, or behavior
tokens~\cite{p5rec,mbgen,gensar2025}. Related approaches use register tokens or
context-dependent beginning-of-sequence (BOS) queries~\cite{yang2025earn,liu2026decor}. When used for
objective differentiation, these mechanisms introduce only a small amount of
objective-specific state, mainly in the input or initial decoding state. The
core transformations that determine token probabilities remain shared. They
therefore change how a common policy is conditioned rather than give each
objective dedicated trainable transformations. The objective policies remain
coupled during optimization. At inference, separate objective-conditioned runs
may also favor the same high-probability Semantic ID regions and produce
overlapping candidate lists under the fixed budget.
Panels (a) and (b) of Figure~\ref{fig:intro_overview} summarize this central
trade-off. Traditional multi-route retrieval is controllable but fragmented,
whereas single-decoder generative retrieval is unified but coupled.

Resolving this trade-off requires satisfying three coupled requirements. \textbf{First}, shared
modeling must support independent objective updates. Reusing user
representations and a common Semantic ID (SID) generation prior avoids full-decoder
duplication, but each objective still needs enough trainable capacity to adapt
without perturbing the general policy or other objectives. Copying a complete
decoder for every objective would restore independence only by recreating
model, training, and serving fragmentation. \textbf{Second}, optimization must
accommodate heterogeneous feedback while preserving generation validity.
Different objectives may require different sample construction and learning
signals, yet specialization must retain the valid SID distribution learned
from broad exposure. \textbf{Third}, specialized policies must contribute complementary
candidates. Independent policies may still produce overlapping
high-probability candidates, so explicit quotas and coordinated search are
needed to prevent overlap from wasting the fixed budget. A practical framework
must therefore be shared in representation, isolated in objective-specific
updates, and coordinated in decoding.

Panel (c) of Figure~\ref{fig:intro_overview} presents our solution,
\textbf{Multi-Decoder OneRec}.
\textbf{(1) Shared yet isolated architecture.} A shared user-context module
provides reusable user representations, while the General Decoder supplies
base weights and a common SID generation prior. Each objective attaches an
isolated expert containing Low-Rank Adaptation (LoRA)~\cite{hu2022lora} updates, an
objective-specific BOS embedding, and an additive offset to the shared SID
embedding table.
Together with the General Decoder, the expert forms an objective-specific decoder.
\textbf{(2) Feedback-adaptive optimization with gradient isolation.} The
General Decoder and objective experts are trained concurrently from separate
supervision streams with disjoint gradient paths. Exposure-sample NTP updates
the shared base, while each objective loss updates only its corresponding
expert state, with gradients stopped at the shared base parameters. Event-based
experts use NTP on objective-filtered samples, whereas the Watch-time expert
uses policy optimization with a user-history-normalized relative
reward~\cite{deng2025onerec,onerecv2,shao2024deepseekmath}. KL regularization
uses the stop-gradient distribution of the General Decoder as the reference,
anchoring the Watch-time expert to the common SID prior while the General
Decoder continues learning from exposure samples. \textbf{(3) Quota-aware
coordinated decoding.} Multi-Decoder Constrained Beam Search (MD-CBS) assigns
each objective-specific decoder an explicit quota and masks SID prefixes claimed by
earlier routes at a configurable decoding
level~\cite{vijayakumar2018diverse,anderson2017constrained,
post2018constrained}. The constraint level determines whether exact SIDs or
broader semantic regions are excluded. The General Decoder runs last and
backfills any residual budget with legal candidates not selected earlier.

We evaluate Multi-Decoder OneRec offline and in production. For reproducible
offline evaluation, we publicly release
\textbf{Kwai26}, a large-scale benchmark for multi-objective generative
retrieval with 1.31 billion raw item-level records, 31.85 million Item-ID
entries, and 25.03 million items with valid Semantic IDs. Under the same
512-item retrieval budget, Multi-Decoder OneRec achieves relative gains of
1.69\%--5.62\% over the single-decoder OneRec~\cite{deng2025onerec} baseline across four Recall@512
metrics. In the production A/B test, it yields relative gains of 0.37\% in app
usage time per device, 0.19\% in Day-7 retained users, 0.19\% in devices with at
least one share, and 2.09\% in Cold-Start. Together, the offline and
online results show that Multi-Decoder OneRec improves specialized objectives
without sacrificing general exposure retrieval and that these gains translate
into measurable production value. Our main contributions are:

\begin{itemize}[leftmargin=15pt, topsep=2pt, itemsep=1pt, parsep=0pt,
  partopsep=0pt]
  \item To the best of our knowledge, \textbf{Multi-Decoder OneRec} is the
  first framework to equip a shared SID retriever with objective-specific
  objective-specific decoders, gradient-isolated adaptation, explicit per-route quotas,
  and cross-decoder constrained search under a fixed candidate budget.
  \item We develop expert optimization with isolated gradient paths and
  Multi-Decoder Constrained Beam Search (MD-CBS). NTP on objective-filtered
  samples and policy optimization with relative rewards address event-based
  and continuous feedback, respectively, while reference regularization
  preserves the general SID prior. MD-CBS combines explicit route quotas with
  prefix constraints across routes to reduce overlap and improve candidate
  complementarity.
  \item We publicly release \textbf{Kwai26}, a billion-scale benchmark for
  multi-objective generative retrieval, together with predefined splits and an
  evaluation protocol. Extensive offline experiments and a production A/B test
  validate Multi-Decoder OneRec at industrial scale.
\end{itemize}

\section{Related Work}
\label{sec:related_work}

\subsection{Generative Recommendation}

\runinhead{Semantic ID Modeling}
Generative retrieval produces Semantic ID (SID) sequences instead of scoring a
fixed catalog as in SASRec~\cite{kang2018sasrec} and
BERT4Rec~\cite{sun2019bert4rec}. TIGER~\cite{rajput2023tiger} introduced
hierarchical SIDs; later work explores self-supervision
\cite{jin2024lmindexer}, multi-signal learning~\cite{wang2024letter},
contextual tokenization or representation~\cite{hou2025actionpiece,
liu2026decor}, and joint tokenizer--recommender optimization
\cite{bai2026bloger}. RPG~\cite{hou2025rpg} and
DiffGRM~\cite{liu2026diffgrm} improve decoding through parallel generation and
discrete diffusion, respectively. These methods focus on one generator rather
than fixed-budget, gradient-isolated objective policies.

\runinhead{Preference Alignment}
Standard NTP does not encode graded feedback or downstream utility.
OneRec~\cite{deng2025onerec,onerec2025technical} and
OneRec-V2~\cite{onerecv2} introduce iterative alignment, duration-aware
rewards, and group-based optimization. GenRec~\cite{zou2026genrec} combines
page-wise NTP, GRPO-SR, and hybrid rewards; GFlowGR~\cite{wang2026gflowgr}
optimizes generation trajectories; and UGR~\cite{fan2026ugr} models reward
uncertainty. All generally align one policy, leaving isolated updates across
concurrent objectives unexplored.

\subsection{Multi-Objective Retrieval}

\runinhead{Objective-Specific Adaptation}
P5~\cite{p5rec}, MBGen~\cite{mbgen}, and GenSAR~\cite{gensar2025} condition
shared generators with prompts or behavior tokens, but do not provide
objective-specific transformations or gradient isolation.
PinRec~\cite{botta2026pinrec} conditions a unified model on
surface-specific outcomes, while EAGER~\cite{wang2024eager} separates
behavioral and semantic token streams. MMoE~\cite{mmoe} and PLE~\cite{ple}
motivate shared--specific decomposition, and LoRA~\cite{hu2022lora} enables
parameter-efficient adaptation.

\runinhead{Quota-Aware Decoding}
Multi-route retrieval studies route fusion and trigger selection
\cite{xie2020ican,nie2022mic,huang2025multichannel,zhou2026capts}, retaining
quota control through route-specific models or serving logic. Diverse beam
search~\cite{vijayakumar2018diverse} diversifies one decoder, while constrained
decoding~\cite{anderson2017constrained,post2018constrained} restricts tokens or
prefixes; both face known limitations in generative retrieval
\cite{wu2025constrained}. Prior work does not coordinate quotas across
objective-specific generators in one SID space. Multi-Decoder OneRec combines
a shared SID prior, isolated adaptation, and quota-aware cross-route decoding.

\section{Problem Formulation}
\label{sec:problem_formulation}

Let $\mathcal{U}$ and $\mathcal{I}$ be the user and item sets, and
$\mathcal{I}_{\mathrm{SID}}\subseteq\mathcal{I}$ the retrievable items with
valid Semantic IDs. At user $u$'s request time $\tau$, the session is the
ordered served-item list and the available history is strictly time-truncated:
$H_u^\tau=(x_1,\ldots,x_n)$. Each pre-$\tau$ interaction records an Item-ID,
content and author features, feedback, and a timestamp. Following generative
retrieval~\cite{rajput2023tiger,deng2025onerec}, item
$i\in\mathcal{I}_{\mathrm{SID}}$ has code
\begin{equation}
  \mathbf{z}(i)=(z_0(i),\ldots,z_{L-1}(i)), \qquad
  z_\ell(i)\in\mathcal{V}_\ell,\quad 0\le\ell<L,
\end{equation}
where $\mathcal{V}_\ell$ is the level-$\ell$ vocabulary. Routes share the
canonical code-to-item mapping. Let $\mathcal{R}=\{\mathrm{gen}\}\cup\mathcal{T}$
with objective routes $\mathcal{T}=\{1,\ldots,M\}$. Route $r$ adds a residual
to the General Decoder's base embedding $E_{\mathrm{gen},\ell}$:
\begin{equation}
  \mathbf{e}_{r,\ell}(z)
  =\bigl(E_{\mathrm{gen},\ell}+\Delta E_{r,\ell}\bigr)[z],
  \qquad r\in\mathcal{R},
  \label{eq:branch_lookup}
\end{equation}
where $\Delta E_{\mathrm{gen},\ell}=0$; item identity is shared, but its
decoding representation is route-adaptive.

Route $\mathrm{gen}$ performs exposure retrieval. The specialized objectives
are Long-View, Like, and Watch-time offline, and Watch-time, Share, and
Cold-Start in production; only their labels and supervision differ. Each route
autoregressively produces ordered list $R_r(u)$ via
\begin{equation}
  p_r(\mathbf{z}\mid H_u^\tau)
  =\prod_{\ell=0}^{L-1}
   p_r(z_\ell\mid z_{<\ell},H_u^\tau),
  \label{eq:branch_policy}
\end{equation}
where $z_{<\ell}$ is the generated prefix. Quotas satisfy
$q_{\mathrm{gen}}+\sum_{t\in\mathcal{T}}q_t=B$ and yield candidate pool
\begin{equation}
  C(u;B)=\operatorname{Merge}_{\pi}
  \left(\{(R_r(u),q_r)\}_{r\in\mathcal{R}}\right),
  \qquad |C(u;B)|\le B,
  \label{eq:quota}
\end{equation}
where $\operatorname{Merge}_{\pi}$ follows route order $\pi$, removes items
accepted earlier, and retains the current route's highest-probability legal
items. Before merging, constrained decoding prevents later routes from
regenerating selected prefixes; the General Decoder runs last and retrieves up
to its quota of unseen legal items. Uncoordinated overlap loses capacity
according to
\begin{equation}
  \mathrm{Dup}(u)
  =1-\frac{\left|\bigcup_{r\in\mathcal{R}}C_r(u;q_r)\right|}
  {\sum_{r\in\mathcal{R}}|C_r(u;q_r)|},
  \label{eq:duplicate}
\end{equation}
where $C_r(u;q_r)$ is route $r$'s quota-truncated, pre-merge output.

For discrete objective $a$ with held-out targets $Y_a(u)$, item-level recall is
\begin{equation}
  \mathrm{Recall}^{(a)}@B
  =\frac{\sum_u |C(u;B)\cap Y_a(u)|}
  {\sum_u |Y_a(u)|}.
  \label{eq:recall}
\end{equation}
For Watch-time, weighted recall is
\begin{equation}
  \mathrm{WTRecall}@B
  =\frac{\sum_u\sum_{i\in Y(u)}
    w_{u,i}\mathbb{I}[i\in C(u;B)]}
  {\sum_u\sum_{i\in Y(u)}w_{u,i}},
  \label{eq:wt_recall}
\end{equation}
where $Y(u)$ contains all held-out targets, $w_{u,i}$ is observed watch time,
and $\mathbb{I}[\cdot]$ is the indicator. We seek higher general and
objective-specific recall under budget $B$, bounded task parameters and
decoding cost, and low duplication. LoRA-expert-$t$ gradients cannot enter the
shared base or other experts, while base NTP continues updating shared
parameters.

\section{Multi-Decoder OneRec}
\label{sec:method}

\begin{figure*}[t]
  \centering
  \includegraphics[width=\textwidth]{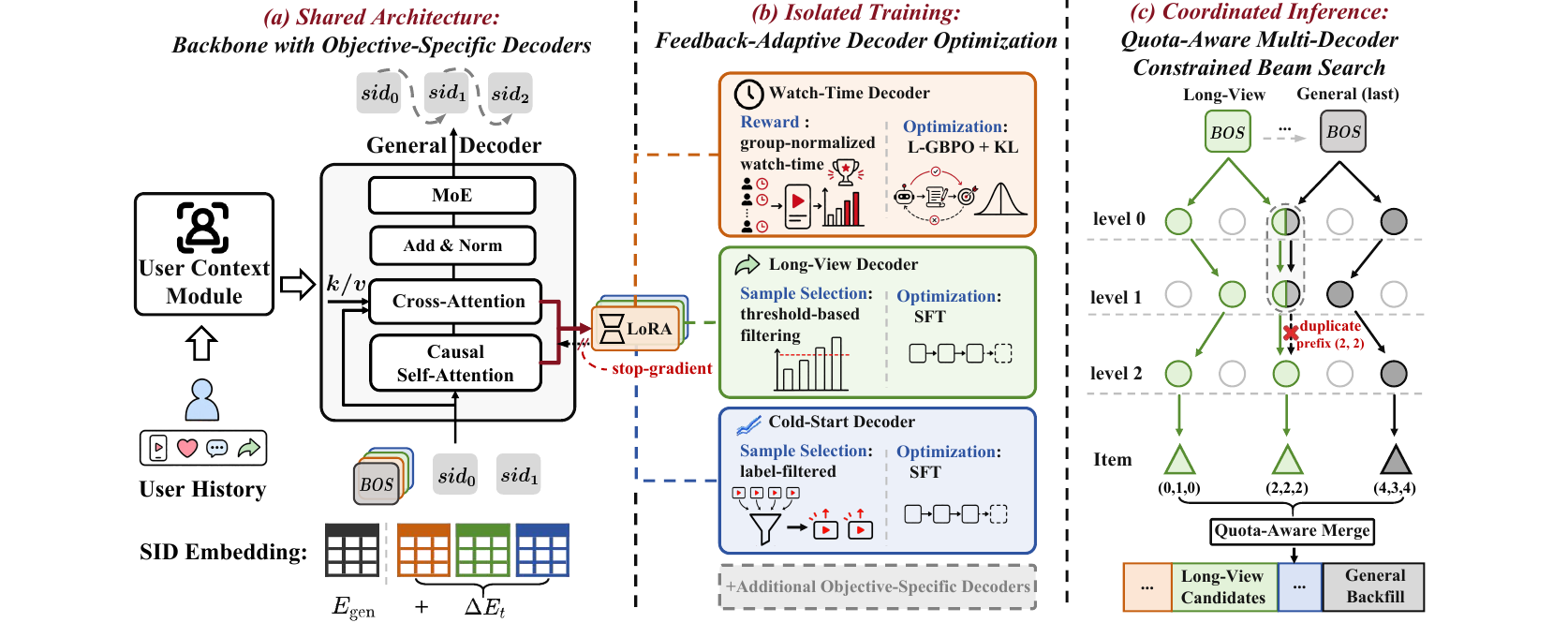}
  \vspace{-10pt}
  \caption{Overview of Multi-Decoder OneRec. (a) \textbf{Shared architecture:}
  user history is encoded once by the user-context module and reused by a
  fully parameterized General Decoder and all objective-specific
  decoders. Each objective reuses the shared backbone while adding an isolated
  BOS embedding, SID embedding residual $\Delta E_t$, and LoRA expert; its
  gradients are stopped at the shared parameters. (b) \textbf{Isolated
  training:} each decoder uses objective-appropriate supervision. The
  Watch-time Decoder uses group-normalized watch-time rewards with L-GBPO and KL
  regularization, whereas Long-View, Cold-Start, and other discrete-feedback
  decoders use filtered samples with SFT. (c) \textbf{Coordinated inference:}
  quota-aware MD-CBS executes routes by priority, masks SID prefixes already
  claimed by earlier routes to prevent duplicate beam expansion, and runs the
  General Decoder last to backfill the fixed-budget candidate pool before
  quota-aware merging.}
  \Description{Multi-Decoder OneRec has three stages. Panel (a) shows user
  history entering a shared user-context module and General Decoder; each
  objective-specific decoder adds an isolated BOS embedding, SID residual, and
  LoRA expert with stopped gradients to the shared base. Panel (b) shows a
  Watch-time decoder trained with normalized rewards, L-GBPO, and KL regularization,
  and Long-View and Cold-Start decoders trained by filtered-sample SFT. Panel (c)
  shows priority-ordered constrained beam search: later routes reject SID
  prefixes selected by earlier routes, and the General Decoder fills the
  remaining quota before all candidates are merged.}
  \label{fig:method_framework}
  \vspace{-15pt}
\end{figure*}

As shown in Figure~\ref{fig:method_framework}, Multi-Decoder OneRec builds multiple objective-specific decoders on a shared
OneRec-style encoder--decoder backbone
~\cite{deng2025onerec,onerec2025technical}. It contains a user-context module,
a fully parameterized General Decoder, and $M$ objective-specific
decoders. The General Decoder learns general exposure retrieval and provides
the common SID generation prior. Objective LoRA expert
$t$ reuses the base weights and adds only a task BOS embedding, a SID embedding
residual, and LoRA parameters. Shared parameters capture transferable user
interests and SID syntax. Isolated task parameters support gradient-isolated
objective updates. Coordinated decoding turns the policy differences into
complementary candidates under a fixed budget. During concurrent optimization,
the exposure loss updates the shared user-context module and General Decoder.
Each objective loss stops at these shared parameters and updates only its
corresponding expert, leaving every other expert unchanged.

\subsection{User-Context Prefilling}

As illustrated in Figure~\ref{fig:method_framework}(a), on the encoder side,
each request is represented by two time-truncated behavior
views: the 20 most recent interacted videos and 256 Long-View videos from the
user's historical sequence. This design lets the model jointly capture
short-term intent and long-term preference before any objective-specific
decoding is performed. For each video, we construct a five-field feature tuple,
\emph{i.e.}, Item-ID, author ID, tag, time difference, and watch time. The
fields are first mapped into embeddings and then processed by a four-block
encoder. Each encoder block contains feature-level cross-attention and
sequence-level self-attention. The cross-attention sublayer fuses heterogeneous
item features, while the self-attention sublayer models dependencies inside
the user's behavior sequence. After stacked encoding, the model obtains the
user-context states
$\mathbf{H}_u=\operatorname{Enc}(H_u^\tau)$, which are computed once per
request and reused as the keys and values of the cross-attention modules in all
decoder routes.

\subsection{Multi-Decoder}

Figure~\ref{fig:method_framework}(a) also shows the shared General Decoder and
its lightweight objective-specific branches. On the decoder side, each target
video is represented by a sequence of discrete Semantic IDs (SIDs). Each SID
token is mapped to a vector through an embedding lookup table. During training,
the input to the General Decoder
(Decoder$_0$) is formed by concatenating a learnable beginning-of-sequence
(BOS) embedding with the right-shifted target SID sequence. The resulting
sequence is processed by stacked Transformer decoder blocks, each consisting
of masked self-attention, cross-attention, and a feed-forward network (FFN).
Masked self-attention preserves the autoregressive factorization of SID
generation, cross-attention injects the encoded user-interest context
$\mathbf{H}_u$, and the FFN further transforms the token representations
nonlinearly.

To adapt the shared generation backbone to different downstream objectives without duplicating the full decoder, we introduce an independent set of LoRA parameters for each task on top of Decoder$_0$. In industrial recommender systems, maintaining a full decoder for every task would incur prohibitive computational, memory, and deployment costs as the number of tasks grows. Moreover, many downstream objectives have highly sparse supervision, making it difficult to adequately train a large task-specific decoder and leading to poorly estimated parameters and suboptimal generalization. LoRA instead enables parameter-efficient task specialization while preserving the knowledge shared through the backbone. Specifically, LoRA updates are
attached to the query, key, and value projection matrices in both masked
self-attention and cross-attention. For task $t$, each adapted projection is
written as:
\begin{equation}
  \mathbf{W}^{(t)}=\mathbf{W}_0+\Delta\mathbf{W}^{(t)}
  =\mathbf{W}_0+\mathbf{B}^{(t)}\mathbf{A}^{(t)},
  \label{eq:lora}
\end{equation}
where $\mathbf{W}_0$ is the shared projection matrix inherited from Decoder$_0$,
$\Delta\mathbf{W}^{(t)}$ is the task-specific low-rank update, and
$\mathbf{A}^{(t)}\in\mathbb{R}^{r_t\times d_{\mathrm{in}}}$ and
$\mathbf{B}^{(t)}\in\mathbb{R}^{d_{\mathrm{out}}\times r_t}$ are independently learned
low-rank matrices for task $t$; $d_{\mathrm{in}}$ and $d_{\mathrm{out}}$ are
the projection's input and output dimensions, with
$r_t\ll\min(d_{\mathrm{in}},d_{\mathrm{out}})$. These low-rank increments let
each task learn its own representation and generation pattern while reusing the
general SID-generation capability of the shared decoder.

In addition, every downstream task owns an independent learnable BOS embedding
$\mathbf{b}_t$
to explicitly inject task information at the start of generation. Its SID
lookup table is also composed of the shared Decoder$_0$ embedding and a
task-specific residual, as defined in Eq.~\eqref{eq:branch_lookup}. We collect
the complete trainable state of task $t$ as:
\begin{equation}
  \phi_t=\{\mathbf{A}^{(t)},\mathbf{B}^{(t)},\mathbf{b}_t,\Delta E_t\},
\end{equation}
where $\mathbf{A}^{(t)}$ and $\mathbf{B}^{(t)}$ denote all low-rank matrices
attached to the attention projections and
$\Delta E_t=\{\Delta E_{t,\ell}\}_{\ell=0}^{L-1}$ collects the SID embedding
residuals across levels. The state $\phi_t$ is isolated from all other experts,
so updating one objective does not alter the General Decoder or another
LoRA expert.

\subsection{Objective-Specific Training}
\label{sec:objective_training}
As summarized in Figure~\ref{fig:method_framework}(b), the supervision samples
assigned to a decoder and its loss jointly shape decoder behavior. Sample
selection defines the target behavior, while the objective
determines how the decoder updates its generation policy toward that behavior.
To prevent newly introduced objectives from interfering with Decoder$_0$ or
with one another, gradients from every objective loss are prevented from
propagating into the shared user-context module and General Decoder; the exposure loss
continues to update both shared components.
\runinhead{Next-Token Prediction}
The General Decoder is trained on exposed items using autoregressive next-token
prediction. Given user context $c$ and target SID
$\mathbf{y}=\mathbf{z}(i)=(y_0,\ldots,y_{L-1})$, we write the base loss as:
\begin{equation}
  \mathcal{L}_{\mathrm{base}}
  =-\sum_{\ell=0}^{L-1}
  \log p_{\mathrm{gen}}(y_\ell\mid c,y_{<\ell}),
  \label{eq:base_sft}
\end{equation} 
Moreover, discrete-task LoRA
experts are trained only on samples satisfying the corresponding behavior
condition. The Like decoder uses liked items, and the Long-View decoder uses
Long-View items. We write the task supervised fine-tuning (SFT) loss as:
\begin{equation}
  \mathcal{L}_{\mathrm{SFT}}^{(t)}
  =-\sum_i\sum_{\ell=0}^{L-1}m_i^{(t)}
  \log p_t(y_{i,\ell}\mid c_i,y_{i,<\ell}),
  \label{eq:sft}
\end{equation}
where $m_i^{(t)}$ indicates whether sample $i$ is valid for objective $t$.
For this objective stream, only $\phi_t$ receives gradients, so its update
changes neither the shared base nor another expert.

\runinhead{Reward-Based Policy Optimization}
For the continuous Watch-time objective considered here, binarized NTP would
provide only coarse supervision because thresholding collapses preference
magnitudes among positive samples. We instead construct a personalized relative
reward that preserves the strength of continuous feedback.
For sample $i$, let $\mathcal{G}_i$ contain its watch time and those of the
user's latest $K$ valid historical items. We write the standardized reward as:
\begin{equation}
  r_i=\frac{w_i-\mu(\mathcal{G}_i)}
  {\sigma(\mathcal{G}_i)+\epsilon},
  \label{eq:relative_reward}
\end{equation}
where $w_i$ is the current watch time (the sample-indexed form of $w_{u,i}$),
$\mu$ and $\sigma$ are group statistics, and $\epsilon>0$ is a numerical
stabilizer. This adapts group-relative reward normalization
~\cite{shao2024deepseekmath,onerecv2} to a user's logged history. The task
decoder is the policy $p_t$, and a stop-gradient forward pass through the
current General Decoder supplies the reference probability $p_{\mathrm{ref}}$.
For compactness, define the ground-truth token probabilities
$p_{t,i,\ell}=p_t(y_{i,\ell}\mid c_i,y_{i,<\ell})$ and
$p_{\mathrm{ref},i,\ell}=p_{\mathrm{ref}}(y_{i,\ell}\mid
c_i,y_{i,<\ell})$. We write the stop-gradient denominator as:
\begin{equation}
  p_{\mathrm{old},i,\ell}^{(t)}=
  \begin{cases}
    \max\!\left(p_{\mathrm{ref},i,\ell},
      \operatorname{sg}(p_{t,i,\ell})\right), & r_i\ge 0,\\
    \max\!\left(p_{\mathrm{ref},i,\ell},
      1-\operatorname{sg}(p_{t,i,\ell})\right), & r_i<0
  \end{cases},
  \label{eq:p_old}
\end{equation}
where $\operatorname{sg}(\cdot)$ stops gradients. We write the life-long gradient-bounded policy optimization (L-GBPO) objective as:
\begin{equation}
  \mathcal{L}_{\mathrm{L-GBPO}}^{(t)}
  =-\mathbb{E}_{i,\ell}\left[
  r_i\frac{p_{t,i,\ell}}
  {p_{\mathrm{old},i,\ell}^{(t)}+\epsilon}
  \right],
  \label{eq:gbpo}
\end{equation}
where the sample-level reward $r_i$ is shared by all SID positions of item $i$.
Reference-policy regularization discourages preference optimization from
drifting away from the general SID prior~\cite{ouyang2022instructgpt}. We
define the token probability ratio as:
\begin{equation}
  \rho_{i,\ell}^{(t)}=
  \frac{p_{\mathrm{ref},i,\ell}}
  {p_{t,i,\ell}},
\end{equation}
where $\rho_{i,\ell}^{(t)}$ compares the reference and task policies on the observed
token. We write the non-negative observed-token KL surrogate
~\cite{shao2024deepseekmath} as:
\begin{equation}
  \mathcal{L}_{\mathrm{KL}}^{(t)}
  =\mathbb{E}_{i,\ell}
  \left[\rho_{i,\ell}^{(t)}-\log \rho_{i,\ell}^{(t)}-1\right],
  \label{eq:kl}
\end{equation}
where the expectation averages over valid logged training positions. We write the objective
for the Watch-time expert $t$ as:
\begin{equation}
  \mathcal{L}_{\mathrm{RL}}^{(t)}
  =\lambda_{\mathrm{L-GBPO}}^{(t)}\mathcal{L}_{\mathrm{L-GBPO}}^{(t)}
  +\lambda_{\mathrm{KL}}^{(t)}\mathcal{L}_{\mathrm{KL}}^{(t)},
  \label{eq:rl}
\end{equation}
where $\lambda_{\mathrm{L-GBPO}}^{(t)}$ and $\lambda_{\mathrm{KL}}^{(t)}$ control
preference learning and reference regularization.

\runinhead{Concurrent Optimization with Gradient Isolation}
Exposure and objective supervision are optimized concurrently in one training
process; their batches may be interleaved or mixed. Gradient routing, rather
than temporal freezing, isolates their updates. Let $\theta_0$ denote the
parameters shared by the user-context module and General Decoder. We write the
gradient-routed objective as:
\begin{equation}
  \mathcal{L}
  =\mathcal{L}_{\mathrm{base}}(\theta_0)
  +\lambda_{\mathrm{SFT}}\sum_{t\in\mathcal{T}_{\mathrm{SFT}}}
   \mathcal{L}_{\mathrm{SFT}}^{(t)}(\operatorname{sg}(\theta_0),\phi_t)
  +\sum_{t\in\mathcal{T}_{\mathrm{RL}}}
   \mathcal{L}_{\mathrm{RL}}^{(t)}(\operatorname{sg}(\theta_0),\phi_t),
  \label{eq:total_loss}
\end{equation}
where $\operatorname{sg}(\theta_0)$ allows the current shared base to
participate in each objective-specific decoder's forward pass without receiving
its gradient;
$\mathcal{T}_{\mathrm{SFT}}$ and $\mathcal{T}_{\mathrm{RL}}$ partition the
specialized objectives by training type, and $\lambda_{\mathrm{SFT}}$ weights
the SFT losses.
Consequently, base NTP is the only source of updates to $\theta_0$, and each
objective loss updates only its corresponding $\phi_t$.

\subsection{Multi-Decoder Constrained Beam Search}
\label{sec:constrained_beam}

Figure~\ref{fig:method_framework}(c) illustrates the coordinated inference
procedure. Independent decoding followed by post-hoc de-duplication wastes beam capacity:
different decoders may spend their quotas on the same item, leaving the merged
candidate pool below the nominal retrieval budget. We therefore use MD-CBS, a
multi-decoder variant of constrained decoding~\cite{anderson2017constrained,
post2018constrained}. MD-CBS executes decoder routes in a predefined priority
order and lets each later route avoid candidates already claimed by earlier
routes. Each route is specified as $(r,b_r,q_r)$, where $r$ is the decoder
route, $b_r$ is its beam size, and $q_r$ is its output quota. The quotas sum to
the final retrieval budget $B$, and the General Decoder is placed last to
backfill remaining general-exposure slots.

Because Item SIDs are hierarchical, the constraint can be imposed at any
zero-based SID level $0\le d<L$. Let $\mathbf{z}_{0:d}=(z_0,\ldots,z_d)$ be
the corresponding partial hypothesis. MD-CBS uses it as the
de-duplication key. For route $r$, let $\mathcal{K}^{(d)}_{<r}$ denote all
level-$d$ keys accepted by earlier routes under order $\pi$, and let
$s_r(\mathbf{z}_{0:d})$ be its route score. The masked score
is:
\begin{equation}
  s_r'(\mathbf{z}_{0:d})=
  \begin{cases}
    -\infty, & \mathbf{z}_{0:d}\in\mathcal{K}^{(d)}_{<r},\\
    s_r(\mathbf{z}_{0:d}), & \text{otherwise}.
  \end{cases},
  \label{eq:prefix_mask}
\end{equation}
Route $r$ skips masked candidates and continues down its beam list until either
$q_r$ unique candidates are accepted or its beam is exhausted. Accepted keys are
then inserted into $\mathcal{K}^{(d)}_{\le r}$ before the next route starts.

The constraint level controls a quality--efficiency trade-off. A smaller $d$
removes occupied prefixes before deeper expansion and can therefore reduce
decoding cost, but it also narrows search breadth: once a shallow prefix is
claimed, later routes cannot explore other useful items under the same coarse
semantic region. A larger $d$ delays de-duplication and preserves richer
semantic exploration, but requires more decoding. In our default setting, each
item SID has three levels, $(z_0,z_1,z_2)$, and we set $d=2$, so the
complete SID tuple is used as the key. This prevents repeated complete SIDs
without prematurely excluding coarse semantic regions; item-level duplicates
caused by SID-to-item collisions are removed during merging. Earlier-layer
constraints are evaluated as granularity variants in Section~\ref{sec:experiments}.
Finally, the General Decoder retrieves up to its quota of unseen complete SIDs.
This converts
independently trained objective-specific decoders into a quota-controlled, non-duplicate candidate
pool under the fixed budget $B$.

\subsection{Serving Workflow}
\label{sec:training_serving}

All decoders share one materialized encoder representation at serving time.
Each active decoder applies its own BOS, SID residual, LoRA state, beam size, and
quota before constrained merging. The objective set and quota allocation can
therefore change without modifying the General Decoder or another expert. Each added
decoder still incurs decoder computation, which Section~\ref{sec:further_analysis}
reports separately from parameter count.

\section{Experiments}
\label{sec:experiments}

\subsection{Experimental Setup}
\label{sec:experimental_setup}

\runinhead{Kwai26 Dataset\footnote{\url{https://github.com/liuzhao09/Kwai26-Data-Pipeline}}}
\label{sec:dataset}
We construct Kwai26 from 60 consecutive days of interactions on an
industrial short-video platform. Appendix Table~\ref{tab:dataset_stats}
summarizes the dataset: 1.31 billion raw records yield 821.84 million
positive-play interactions and 125.26 million training sessions, with one
held-out session for each of 50,000 users. Appendix~\ref{app:kwai26} details the
source profile, labels, leakage-safe processing, split, features, and
chronological training layout. We publicly release the predefined split and
evaluation protocol with the dataset.

\begin{table}[!tb]
  \centering
  \caption{Overall offline performance on Kwai26 in Recall@512 ($\uparrow$).
  Best results are in \textbf{bold}, second-best results are
  \underline{underlined}, and gains over the strongest baseline are
  statistically significant ($p<0.05$).}
  \label{tab:overall_results}
  \setlength{\tabcolsep}{1.5pt}
  \renewcommand{\arraystretch}{1.10}
  \begin{tabular*}{\columnwidth}{@{\extracolsep{\fill}}l|lcccc@{}}
    \toprule
    \textbf{Paradigm} & \textbf{Model} & \textbf{Exp.} & \textbf{Long-View} &
      \textbf{Like} & \textbf{WT} \\
    \midrule
    \multirow{3}{*}{Discriminative}
      & DSSM~\cite{huang2013dssm} & 0.0046 & 0.0050 & 0.0039 & 0.0050 \\
      & SASRec~\cite{kang2018sasrec} & 0.0310 & 0.0329 & 0.0254 & 0.0329 \\
      & HSTU~\cite{zhai2024hstu} & 0.0342 & 0.0469 & 0.0387 & 0.0506 \\
    \midrule
    \multirow{3}{*}{Generative}
      & TIGER~\cite{rajput2023tiger} & 0.1380 & 0.1613 & 0.1161 & 0.1660 \\
      & OneRec~\cite{deng2025onerec} & \underline{0.1539} & \underline{0.1833} &
      \underline{0.1318} & \underline{0.1923} \\
      & \textbf{Ours} &
      \textbf{0.1565} & \textbf{0.1907} & \textbf{0.1391} &
      \textbf{0.2031} \\
    \bottomrule
  \end{tabular*}
\end{table}

\runinhead{Baselines}
We compare Multi-Decoder OneRec with five representative retrieval models:
\begin{itemize}[leftmargin=15pt,topsep=2pt,itemsep=1pt,parsep=0pt]
  \item \textbf{DSSM}~\cite{huang2013dssm} encodes the user history and items
  with two towers and performs full-catalog approximate nearest-neighbor (ANN) retrieval.
  \item \textbf{SASRec}~\cite{kang2018sasrec} uses a causal Transformer to
  encode the chronological behavior sequence, followed by ANN retrieval.
  \item \textbf{HSTU}~\cite{zhai2024hstu} uses pointwise aggregated attention
  and gated residual updates as an industrial sequential encoder, followed by
  full-catalog ANN retrieval.
  \item \textbf{TIGER}~\cite{rajput2023tiger} encodes item histories with SIDs
  and autoregressively generates target SID sequences.
  \item \textbf{OneRec}~\cite{deng2025onerec} uses the same multi-field encoder
  as our method and a single decoder that matches our General Decoder.
\end{itemize}
All baselines and our method receive the same sessions, chronological split,
and time-truncated recent-20 and Long-View-256 behavior positions. They differ
only in feature representation and retrieval mechanism. DSSM~\cite{huang2013dssm},
SASRec~\cite{kang2018sasrec}, and HSTU~\cite{zhai2024hstu}
are trained on exposure targets and return the ANN Top-512 over the complete
item catalog. TIGER~\cite{rajput2023tiger}, OneRec~\cite{deng2025onerec}, and our method share the SID lexicon and return 512
items by beam search. Specialized feedback trains only the corresponding decoders
in our method. The General Decoder uses the same exposure supervision as the
baselines.

\runinhead{Implementation Details}
For OneRec~\cite{deng2025onerec}, TIGER~\cite{rajput2023tiger}, and our method, the hidden size is 256, the encoder and
decoder contain four Transformer blocks with eight heads, and the SID has three
levels with 8,192 codes per level. Models are trained for one chronological pass
with a global batch size of 1,024. The default LoRA rank is 32. All methods are
evaluated on the same Kwai26 catalog of 25.03 million items. The default beam
sizes for Long-View, Like, Watch-time, and the General Decoder are 86, 171,
256, and 512. Their quotas are 86, 85, 85, and 256 for a total budget of 512.
We apply MD-CBS at the final level of SID to achieve optimal performance.
\runinhead{Metrics}
We report item-level Recall@512 for Exposure (Exp.), Long-View, and Like,
together with Watch-time (WT) Recall@512. Long-View denotes the platform-defined
Long-View label. The first three metrics follow
Eq.~\eqref{eq:recall}. The last weights every target by its observed watch time
as in Eq.~\eqref{eq:wt_recall}. All recalls are micro-averaged over target items
in the 50,000 test sessions. We additionally use SID legal rate, i.e., the
fraction of generated SID candidates that map to valid Item-ID--SID pairs,
total valid candidate count, parameter count, and floating-point operations
(FLOPs) in diagnostic studies.

\subsection{Overall Performance}
\label{sec:main_results}

Table~\ref{tab:overall_results} reports the overall Recall@512 of
Multi-Decoder OneRec and all baselines under the same 512-item retrieval
budget. The comparison shows two clear trends. First, SID-based generative
retrieval substantially outperforms the three embedding-retrieval baselines on
the large and sparse industrial catalog, confirming the benefit of generating
structured item identifiers instead of relying solely on ANN retrieval in the
embedding space. Second, Multi-Decoder OneRec achieves the best result on every
target set, improving the single-decoder OneRec~\cite{deng2025onerec} baseline by 1.69\%, 4.04\%,
5.54\%, and 5.62\% on Exposure, Long-View, Like, and Watch-time Recall,
respectively. These consistent gains indicate that the quota-controlled union
of objective-specific decoders strengthens objective-specific retrieval while
preserving general exposure coverage.

\subsection{Ablation Studies}
\label{sec:ablation}

\runinhead{Single- and Multi-Decoder Training}
Table~\ref{tab:multi_decoder_ablation} disentangles task-specific decoding from
quota-controlled multi-decoder aggregation. A specialized decoder can optimize
its own target: the Long-View and Watch-time decoders attain the strongest
corresponding recalls. Yet each loses ground elsewhere, showing that one
objective policy cannot serve as a balanced retrieval pool. Ours combines
isolated updates with quota-controlled aggregation and is the only configuration
ranked in the top two on all four metrics. Specialization alone is therefore
insufficient: its value emerges when distinct candidate streams are retained and
merged under a shared budget. This gives a stronger multi-objective operating
point than either the shared General Decoder or any single specialized policy.
Moreover, we compare the proposed method with a Multi-BOS baseline, in which each BOS token is assigned to a specific task variant under the same training protocol and quota allocation. The results show that our method
consistently outperforms this variant across all metrics. This suggests that a
task-specific BOS signal alone is insufficient for effective objective
specialization, whereas LoRA-expert adaptation provides a more expressive
mechanism for learning task-aligned generation patterns.
\begin{table}[!tb]
  \centering
  \caption{Comparison with single-objective decoders.
  All values are Recall@512 ($\uparrow$). Best results are in \textbf{bold}, and
  second-best results are \underline{underlined}.}
  \vspace{-10pt}
  \label{tab:multi_decoder_ablation}
  \normalsize
  \setlength{\tabcolsep}{1pt}
  \renewcommand{\arraystretch}{1.06}
  \begin{tabular}{@{}l|cccc@{}}
    \toprule
    \textbf{Configuration} & \textbf{Exp.} & \textbf{Long-View} & \textbf{Like} &
      \textbf{WT} \\
    \midrule
    General Decoder & \underline{0.1539} & 0.1833 & 0.1318 & 0.1923 \\
    Long-View & 0.1403 & \textbf{0.1956} & 0.1315 & 0.2030 \\
    Like & 0.1268 & 0.1663 & \underline{0.1336} & 0.1806 \\
    Watch-time & 0.1219 & 0.1724 & 0.1137 & \textbf{0.2041} \\
    Multi-BOS & 0.1340 & 0.1728 & 0.1243 & 0.1822 \\
    \textbf{Ours} &
      \textbf{0.1565} & \underline{0.1907} & \textbf{0.1391} &
      \underline{0.2031} \\
    \bottomrule
  \end{tabular}
\end{table}

\runinhead{MD-CBS}
Here, CBS denotes Constrained Beam Search. We use L2 CBS and L3 CBS for
MD-CBS with the cross-route constraint applied at the second and third SID
levels, respectively; No CBS denotes independent decoding without cross-route
constraints.
Without MD-CBS, independent routes repeatedly generate the same items; only
207.67 unique candidates remain on average after de-duplication, and recall is
lowest on all four objectives. Both L2 CBS and L3 CBS restore the full 512-item budget
(Table~\ref{tab:cbs_ablation}), confirming that coordinated constraints recover
capacity otherwise spent on overlap. L2 blocks every continuation under an
occupied second-level prefix, so later routes cannot explore distinct items
within that coarse semantic region. L3 masks only complete SIDs. It preserves
more search breadth while removing exact duplicates, achieves the best recall
on every objective, and is therefore the default operating point. Because L2 CBS
and L3 CBS return the same number of candidates, the higher recalls of L3 CBS cannot be explained
by pool size; they show that delaying de-duplication until complete SIDs avoids
prematurely pruning semantically related but distinct items.
\begin{table}[!tb]
  \centering
  \caption{Effect of MD-CBS granularity. L2 CBS and L3 CBS apply CBS at the
  second and third SID levels, respectively; 
  \textbf{\#Cand.} is the number of
  candidates. Best results are in \textbf{bold}, and second-best
  results are \underline{underlined}.}
  \label{tab:cbs_ablation}
  \normalsize
  \setlength{\tabcolsep}{1pt}
  \renewcommand{\arraystretch}{1.06}
  \begin{tabular}{@{}l|ccccc@{}}
    \toprule
    \textbf{Method} & \textbf{Exp.} & \textbf{Long-View} & \textbf{Like} &
      \textbf{WT} & \textbf{\#Cand.} \\
    \midrule
    L3 CBS & \textbf{0.1565} & \textbf{0.1907} &
      \textbf{0.1391} & \textbf{0.2031} & \textbf{512} \\
    L2 CBS & \underline{0.1522} & \underline{0.1827} &
      \underline{0.1350} & \underline{0.1942} & \textbf{512} \\
    No CBS & 0.1258 & 0.1544 & 0.1122 & 0.1647 & 207.67 \\
    \bottomrule
  \end{tabular}
\end{table}

\subsection{A/B Test}
\label{sec:online_ab}

We conduct a seven-day A/B test on the generative recommendation pipeline of
Kwai Brazil. This production pipeline is centered on generative retrieval, whose
retrieved candidates contribute 57\% of total impressions, followed by ranking
and re-ranking stages. Experimental traffic is assigned by device identifier
(DID), which serves as the randomization unit. The control group contains
14.64\% of DIDs and uses single-decoder OneRec~\cite{deng2025onerec}. The
treatment group contains 7.32\% of DIDs, retains the same shared encoder and
General Decoder, and adds Watch-time, Share, and Cold-Start decoders with explicit
route quotas. Apart from the generative retrieval module, both groups use the
same ranking stages.

Under the same downstream stack, the core product metric---usage time per
device---improves by +0.37\%, alongside Day-1/7 retention
(+0.12\%/+0.19\%), interactions (+0.19\%--+0.52\%), and Cold-Start
(+2.09\%) (Table~\ref{tab:online_ab}). The ten outcomes span retention,
consumption, engagement, and ecosystem coverage; Appendix~\ref{app:online_scope}
records the shared reporting scope and the remaining profile-visit outcome.
\begin{table}[!tb]
  \centering
  \caption{Online A/B test performance in short-videos services of Kwai Brazil. (all $p<0.05$).}
  \vspace{-10pt}
  \label{tab:online_ab}
  \normalsize
  \setlength{\tabcolsep}{5pt}
  \renewcommand{\arraystretch}{0.90}
  \begin{tabular}{@{}l|r@{}}
    \toprule
    \textbf{Metric} & \textbf{Change} \\
    \midrule
    \textbf{Usage time per device} & \textbf{+0.37\%} \\
    Day-1 retained users & +0.12\% \\
    Day-3 retained users & +0.15\% \\
    Day-7 retained users & +0.19\% \\
    Share devices & +0.19\% \\
    Like devices & +0.35\% \\
    Comment devices & +0.52\% \\
    Follow devices & +0.51\% \\
    Download devices & +0.50\% \\
    Cold-Start & +2.09\% \\
    \bottomrule
  \end{tabular}
\end{table}

\subsection{Further Analysis}
\label{sec:further_analysis}

\runinhead{Source-Level Online Posterior Analysis}
Appendix Table~\ref{tab:source_posterior} shows that the Share, Cold-Start, and
Watch-time decoders lead in FTR, Cold-Start rate, and average watch time,
respectively. This alignment confirms decoder specialization after downstream
serving.

\begin{figure}[!tb]
  \centering
  \vspace{-15pt}
  \includegraphics[width=\columnwidth]{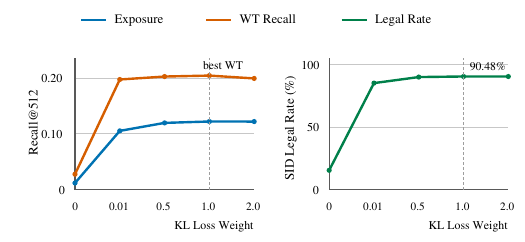}
  \caption{KL-weight sensitivity. Removing KL collapses recall and the SID
  legal rate, while weight 1.0 gives the best Watch-time Recall.}
  \Description{Two plots show exposure and Watch-time Recall and the SID legal
  rate across KL weights from zero to two. All metrics collapse without KL and
  are strongest or stable around weight one.}
  \label{fig:kl_weight_sweep}
\end{figure}

\runinhead{Effect of KL Weight}
Figure~\ref{fig:kl_weight_sweep} studies the sensitivity of the Watch-time RL
expert to the KL weight. A moderate constraint works best: weight 1.0 achieves
the highest Watch-time Recall while preserving the strongest exposure recall
and SID legal rate. Reducing the weight weakens the reference prior and causes a
gradual drop in both the SID legal rate and retrieval quality; removing KL
entirely leads to a severe collapse. These results suggest that the General Decoder
should continue to provide a reference distribution during RL adaptation, rather
than serving only as the initial parameterization.

\runinhead{Reward-History Size}
Using the latest 500 historical items for reward normalization improves WT
Recall from 0.2003 to 0.2031 relative to $K=8$, consistent with more stable
user-relative reward estimation; the complete ablation is reported in
Appendix~\ref{app:hyperparameter_sensitivity}.

\runinhead{Loss Function Selection for Continuous Targets}
Table~\ref{tab:learning_objective} compares two training objectives for the Watch-time expert. For continuous signals such as watch time, threshold-based SFT reduces graded feedback to binary labels, thereby discarding magnitude differences among positive samples. In contrast, continuous-reward RL preserves the relative strength of user preferences and improves WT Recall from 0.1999 to 0.2031. These results demonstrate the advantage of reward-based optimization for modeling fine-grained continuous feedback.
\begin{table}[!tb]
  \centering
  \caption{Effect of learning objective on Watch-time Recall@512. Best result
  is in \textbf{bold}.}
  \label{tab:learning_objective}
  \normalsize
  \setlength{\tabcolsep}{4pt}
  \renewcommand{\arraystretch}{1.06}
  \begin{tabular}{@{}l|c@{}}
    \toprule
    \textbf{Objective} & \textbf{WT Recall} \\
    \midrule
    SFT with 12s threshold & 0.1999 \\
    Continuous-reward RL & \textbf{0.2031} \\
    \bottomrule
  \end{tabular}
  \vspace{-10pt}
\end{table}

\begin{figure}[!tb]
  \centering
  \vspace{-10pt}
  \includegraphics[width=\columnwidth]{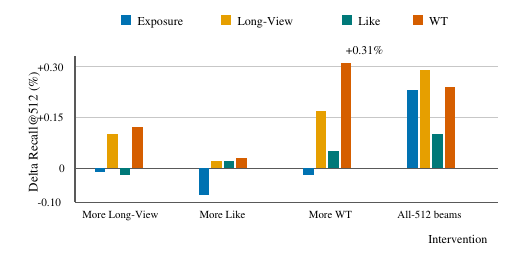}
  \vspace{-10pt}
  \caption{Recall@512 deltas from the default configuration. Colors denote
  target metrics, and x-axis groups denote retrieval interventions.}
  \Description{Grouped bars show changes in four Recall at 512 metrics when
  quota is shifted toward the Long-View, Like, or Watch-time route, or when
  every route beam is enlarged to 512.}
  \label{fig:beam_scaling}
\end{figure}

\runinhead{Quota Allocation and Beam Scaling}
Figure~\ref{fig:beam_scaling} shows Recall@512 deltas from the default setting.
Colors denote Exposure, Long-View, Like, and WT Recall, while x-axis groups denote
retrieval interventions. The first three groups shift quota toward one target.
They steer recall in the intended direction without uniformly improving all
metrics: More Long-View raises Long-View Recall by +0.10\% but slightly lowers Exposure
and Like; More Like gives a small Like gain (+0.02\%); and More WT yields the
largest targeted change, +0.31\% on WT Recall. Because the output budget is
fixed, assigning more quota to one route can remove candidates from others. The
rightmost group, All-512 beams, instead increases beam width and makes every bar
positive, especially on Long-View and WT. Thus, quota allocation controls the
retrieval direction, whereas beam scaling improves candidate quality when extra
compute is available.

\runinhead{LoRA Rank}
We use rank 32, which yields the best result on all four Recall@512 metrics; the
complete ablation and analysis are reported in
Appendix~\ref{app:hyperparameter_sensitivity}.

\runinhead{Efficiency}
Table~\ref{tab:parameter_flops} shows that three LoRA experts add only 20\%
parameters, while training and four-route inference require 1.54$\times$ and
2.23$\times$ FLOPs. Route-specific beams and quotas control this quality--cost
trade-off.
\begin{table}[!tb]
  \centering
  \caption{Parameter and computation overhead. FLOPs are measured per session.}
  \label{tab:parameter_flops}
  \begin{tabular}{lcc}
    \toprule
    Configuration & Params. (rel.) & FLOPs (rel.) \\
    \midrule
    Train: General Decoder & 7.93M (1.00$\times$) & 14.93G (1.00$\times$) \\
    Train: +3 LoRAs        & 9.50M (1.20$\times$) & 22.97G (1.54$\times$) \\
    \midrule
    Infer: General Decoder & 7.93M (1.00$\times$) & 325.02G (1.00$\times$) \\
    Infer: 4 routes        & 9.50M (1.20$\times$) & 725.38G (2.23$\times$) \\
    \bottomrule
  \end{tabular}
\end{table}

\section{Conclusion}
\label{sec:conclusion}

We presented Multi-Decoder OneRec for controllable multi-objective generative
retrieval. First, it combines a shared user-context module and General Decoder
with objective-specific decoders. Second, its
feedback-adaptive optimization uses target-filtered SFT for event-based feedback
and KL-regularized relative-reward policy optimization for the Watch-time decoder, while MD-CBS coordinates explicit route quotas to produce complementary candidates
under a fixed budget. Third, we release Kwai26 with predefined splits and an
evaluation protocol, and validate the framework both offline and in production.
Under the same 512-item budget, Multi-Decoder OneRec improves four Recall@512
metrics by 1.69\%--5.62\% over single-decoder OneRec~\cite{deng2025onerec}; the
production A/B test further improves usage time per device by 0.37\% and
new-content Cold-Start by 2.09\%. These results demonstrate that shared
generative retrieval can retain explicit objective control while producing
complementary candidates at industrial scale.

\bibliographystyle{ACM-Reference-Format}
\bibliography{references}

\appendix
\section{Notation}
\label{app:notation}

Table~\ref{tab:notation} summarizes the symbols used in the formulation and
method.

\begin{table}[!ht]
  \centering
  \caption{Summary of the main notation.}
  \label{tab:notation}
  \footnotesize
  \setlength{\tabcolsep}{4pt}
  \renewcommand{\arraystretch}{1.10}
  \begin{tabular}{@{}>{\centering\arraybackslash}p{0.34\columnwidth}|>{\centering\arraybackslash}p{0.52\columnwidth}@{}}
    \toprule
    \textbf{Symbol} & \textbf{Description} \\
    \midrule
    $\mathcal{U},\mathcal{I},\mathcal{I}_{\mathrm{SID}}$ &
      User, item, and valid-SID item sets \\
    $u,\tau,H_u^\tau$ & User, request time, and truncated history \\
    $L,\mathcal{V}_\ell,\mathbf{z}(i)$ &
      SID level count, level-$\ell$ vocabulary, and item SID \\
    $E_{\mathrm{gen},\ell},\Delta E_{r,\ell}$ &
      General SID embedding and route-$r$ residual \\
    $M,\mathcal{T},\mathcal{R},r,p_r$ &
      LoRA expert count, objective/route sets, route index, and policy \\
    $R_r(u),C_r(u;q_r),q_r$ & Route-$r$ output, truncated output, and quota \\
    $b_r$ & Beam size of route $r$ \\
    $B,\pi,C(u;B)$ & Budget, route order, and candidate pool \\
    $\mathrm{Dup}(u)$ & Cross-route duplication ratio \\
    $Y_a(u),Y(u),w_{u,i}$ & Objective-$a$/all target sets and watch time \\
    $\theta_0$ & Shared context and General Decoder weights \\
    $\phi_t,r_t$ & Expert-$t$ state and LoRA rank \\
    $\mathcal{G}_i,K,r_i$ & Reward group, history size, and relative reward \\
    $\epsilon$ & Positive numerical stabilizer \\
    $p_{t,i,\ell},p_{\mathrm{ref},i,\ell},p_{\mathrm{old},i,\ell}^{(t)}$ &
      Task-decoder, reference, and denominator token probabilities \\
    $\rho_{i,\ell}^{(t)}$ & Reference-to-task-decoder token-probability ratio \\
    $d,\mathcal{K}_{<r}^{(d)}$ &
      Zero-based constraint level and earlier-route SID keys \\
    \bottomrule
  \end{tabular}
\end{table}

\section{Kwai26 Dataset Construction}
\label{app:kwai26}

\begin{table}[!ht]
  \centering
  \caption{Statistics of the Kwai26 offline dataset.}
  \label{tab:dataset_stats}
  \normalsize
  \setlength{\tabcolsep}{6pt}
  \renewcommand{\arraystretch}{1.06}
  \begin{tabular}{@{}c|c@{}}
    \toprule
    \textbf{Statistic} & \textbf{Value} \\
    \midrule
    Users & 50,000 \\
    Raw item-level records & 1,311,923,604 \\
    Positive-play interactions & 821,842,758 \\
    Training sessions & 125,261,311 \\
    Test sessions & 50,000 \\
    Item-ID vocabulary entries & 31,854,181 \\
    Items with valid SIDs & 25,028,687 \\
    SID levels / codes per level & 3 / 8,192 \\
    Recent / Long-View history & 20 / 256 \\
    Maximum targets per session & 8 \\
    \bottomrule
  \end{tabular}
\end{table}

\runinhead{Scope and source profile}
Kwai26 covers 60 consecutive days of short-video traffic from May 2 to
June 30, 2026. Each event records user and request identifiers, event time and
display position, Item-IDs and author IDs, content tags, video and watch times,
behavior flags, and two SID strings. Table~\ref{tab:kwai26_source_profile}
reports source-distribution diagnostics not included in
Table~\ref{tab:dataset_stats}. All are measured before positive-play filtering;
the exposure-rank ranges are disjoint.

\begin{table}[!ht]
  \centering
  \caption{Kwai26 source-distribution diagnostics before positive-play
  filtering.}
  \label{tab:kwai26_source_profile}
  \small
  \setlength{\tabcolsep}{3pt}
  \renewcommand{\arraystretch}{1.06}
  \begin{tabular}{@{}c|c@{}}
    \toprule
    \textbf{Diagnostic} & \textbf{Value} \\
    \midrule
    Cold-Start prevalence & 23.79\% \\
    Effective-view prevalence & 18.66\% \\
    Long-View prevalence & 11.13\% \\
    Like / Share prevalence & 3.22\% / 0.353\% \\
    Single-exposure items & 9,500,929 (29.83\%) \\
    Event share, ranks 1--100 & 0.35\% \\
    Event share, ranks 101--1,000 & 1.07\% \\
    Event share, ranks 1,001--10,000 & 4.62\% \\
    Within-request duplicate rate & 0.0012\% \\
    Events in the first five positions & 69.70\% \\
    \bottomrule
  \end{tabular}
\end{table}

\runinhead{Labels and task views}
Watch-time is a continuous utility signal; Effective View and Long-View are
platform-defined consumption labels; Like and Forward are explicit feedback
signals; and the Cold-Start flag marks new content. Evaluation uses General,
Long-View, Like, and Watch-time views, with the last weighting targets by raw
watch time. Effective View and Cold-Start flags are retained for analysis.

\runinhead{Session definition and canonical order}
A session groups events with the same user and request identifier. Events are stably ordered by time, display position, and Item-ID
without de-duplication; the same order defines targets and user histories.

\begin{algorithm}[!ht]
  \caption{Stage 1: event filtering and materialization}
  \label{alg:kwai26_stage1}
  \small
  \begin{algorithmic}[1]
    \Require Raw event table $\mathcal{D}$
    \Ensure Session table $\mathcal{B}$ and user histories $\{H_u\}$
    \State $\mathcal{E}\gets\emptyset$
    \ForAll{$e\in\mathcal{D}$}
      \State $e\gets\Call{NormalizeMissing}{e}$
      \If{$e.\mathrm{playing\_time}>0$}
        \State $e.\mathrm{reject}\gets\Call{InvalidTargetFields}{e}$
        \State append $e$ to $\mathcal{E}$
      \EndIf
    \EndFor
    \State $\mathcal{B}\gets\Call{StableGroup}{\mathcal{E},
      (\mathrm{user},\mathrm{time},\mathrm{position},\mathrm{item})}$
    \State $\{H_u\}\gets\Call{StableGroup}{\mathcal{E},
      (\mathrm{user},\mathrm{time},\mathrm{position},\mathrm{item})}$
    \State \Return $\mathcal{B},\{H_u\}$
  \end{algorithmic}
\end{algorithm}

\runinhead{Stage 1 details}
Algorithm~\ref{alg:kwai26_stage1} maps missing integer, binary, and string
fields to $-1$, $0$, and the empty string. It removes 490,080,846 non-positive
watch-time events before aggregation; the retained total is reported in
Table~\ref{tab:dataset_stats}. A remaining event is ineligible as a target when
its main SID is missing or an item, duration, timestamp, or request field is
invalid, but it remains available as history. Missing main SIDs account for
58,324,134 of 58,324,137 such events. Stage 1 materializes 126,246,047 sessions
and a chronological history for each user.

\begin{algorithm}[!t]
  \caption{Stage 2: leakage-safe histories, targets, and split}
  \label{alg:kwai26_stage2}
  \small
  \begin{algorithmic}[1]
    \Require Sessions $\mathcal{B}$ and chronological histories $\{H_u\}$
    \Ensure Training set $\mathcal{S}_{\mathrm{tr}}$ and test set
      $\mathcal{S}_{\mathrm{te}}$
    \State $\mathcal{S}_{\mathrm{tr}},\mathcal{S}_{\mathrm{te}}\gets\emptyset$
    \ForAll{users $u$}
      \State $\mathcal{S}_u^{\mathrm{valid}}\gets\emptyset$
      \ForAll{$s\in\Call{Sort}{\mathcal{B}_u,(\mathrm{time})}$}
        \State $\tau\gets s.\mathrm{start};\quad
          j\gets\Call{LowerBound}{H_u.\mathrm{time},\tau}$
        \State $H^{<\tau}\gets H_u[0{:}j]$
        \State $H_{\mathrm{recent}}\gets\Call{Tail}{H^{<\tau},20}$
        \State $H_{\mathrm{LV}}\gets
          \Call{Tail}{\mathrm{FilterLongView}(H^{<\tau}),256}$
        \State $H_{\mathrm{WT}}\gets
          \Call{Tail}{\mathrm{PositivePlay}(H^{<\tau}),500}$
        \State $Y_s\gets\Call{FirstEligible}{s,8}$
        \If{$(H_{\mathrm{recent}}\neq\emptyset\lor
          H_{\mathrm{LV}}\neq\emptyset)\land Y_s\neq\emptyset$}
          \State append $\Call{Encode}{H_{\mathrm{recent}},H_{\mathrm{LV}},
            H_{\mathrm{WT}},Y_s}$ to $\mathcal{S}_u^{\mathrm{valid}}$
        \EndIf
      \EndFor
      \State $\mathcal{S}_{\mathrm{tr}}\gets\mathcal{S}_{\mathrm{tr}}
        \cup\mathcal{S}_u^{\mathrm{valid}}[{:}{-}1]$
      \State $\mathcal{S}_{\mathrm{te}}\gets\mathcal{S}_{\mathrm{te}}
        \cup\{\mathcal{S}_u^{\mathrm{valid}}[-1]\}$
    \EndFor
    \State \Return $\mathcal{S}_{\mathrm{tr}},\mathcal{S}_{\mathrm{te}}$
  \end{algorithmic}
\end{algorithm}

\runinhead{Stage 2 details}
The left-bound \textsc{LowerBound} ensures that every history event precedes
$\tau$; $H_{\mathrm{recent}}$, $H_{\mathrm{LV}}$, and $H_{\mathrm{WT}}$ denote
recent, Long-View-positive, and raw watch-time histories. History positions
contain item, author, tag, bucketed watch time and time gap, mask, and
three-level SID features, and are left-padded. \textsc{FirstEligible} keeps
non-rejected items with a valid three-level SID; targets store an Item-ID and its SID, raw
watch time, and five behavior flags, and are right-padded. Stage 2 removes
49,749 sessions without history and 884,987 without an eligible target.

\runinhead{Split and strict training layout}
After both checks, each user's last valid session is test data and all earlier
sessions are training data; no validation split or date/percentage alternative
is used. Stage 2.5 stably merges training rows by request time, user ID,
request identifier, and sample key without filtering or re-sampling. Each
2,048-row physical block in 32 lanes yields two 1,024-row optimizer steps
(122,324 total); the final 1,535-row tail is not optimized.

\runinhead{Feature vocabularies and SID catalog}
OneRec~\cite{deng2025onerec} uses item, author, tag, watch-time, time-gap, and mask
features; TIGER~\cite{rajput2023tiger} uses SIDs at these positions. The three-level SIDs are generated by
residual-quantization K-Means (RQ-KMeans). The author/tag vocabularies contain 1,829,583/36 entries; the catalog
preserves 27,139,853 valid Item-ID--SID pairs and 23,909,817 codes. Of these,
1,983,033 items have multiple SIDs (at most eight each).

\section{Reproducibility Details}
\label{app:reproducibility}

\begin{table}[!ht]
  \centering
  \caption{Additional training hyperparameters.}
  \label{tab:appendix_hparams}
  \small
  \setlength{\tabcolsep}{8pt}
  \renewcommand{\arraystretch}{1.10}
  \begin{tabular}{@{}c|c@{}}
    \toprule
    \textbf{Hyperparameter} & \textbf{Value} \\
    \midrule
    FFN size & 1,024 \\
    Chronological optimizer steps & 122,324 \\
    Learning rate & $5\times10^{-4}$ \\
    Warm-up steps & 1,000 \\
    Weight decay & 0.15 \\
    \bottomrule
  \end{tabular}
\end{table}

\section{Online Reporting Scope}
\label{app:online_scope}

\begin{table}[!ht]
  \centering
  \caption{Source-level post-serving online diagnostics. FTR denotes the
  share rate, and Avg. Watch-time is measured per video.
  Best results are in \textbf{bold}.}
  \label{tab:source_posterior}
  \normalsize
  \setlength{\tabcolsep}{2.5pt}
  \renewcommand{\arraystretch}{1.06}
  \begin{tabular}{@{}l|ccc@{}}
    \toprule
    \textbf{Source} & \textbf{Cold-Start Rate} & \textbf{FTR} &
      \textbf{Avg. Watch-time} \\
    \midrule
    General Decoder & 0.2079 & 0.0046 & 22.5378 \\
    Share Decoder & 0.0105 & \textbf{0.0077} & 11.9434 \\
    Cold-Start Decoder & \textbf{0.9548} & 0.0026 & 10.4500 \\
    Watch-time Decoder & 0.1960 & 0.0031 & \textbf{26.3564} \\
    \bottomrule
  \end{tabular}
\end{table}

The A/B test reports 11 outcomes. Table~\ref{tab:online_ab} lists ten core
outcomes; profile-visit devices improve by +0.49\% ($p<0.05$). Share, Like,
comment, follow, download, and profile-page outcomes are device counts, and
every reported change is significant at $p<0.05$.

The evaluation covers 1-, 3-, and 7-day retention, app usage time per device,
and key interaction-device counts. Every outcome improves, indicating that
complementary retrieved candidates survive downstream ranking and translate into
consumption, engagement, and ecosystem gains.

\section{Extended Experimental Results}
\label{app:extended_results}

\subsection{Hyperparameter Sensitivity}
\label{app:hyperparameter_sensitivity}

\runinhead{LoRA Rank}
Table~\ref{tab:lora_rank} varies the dimensionality of task-specific updates.
The results are not monotonic in rank: rank 4 and rank 64 provide the
second-best results on different metrics, while rank 16 trails both on all four.
Rank 32 is best on every metric, and increasing the rank to 64 adds parameters
without improving recall. We therefore use rank 32 as the default
adaptation-capacity/parameter-efficiency trade-off.

\begin{table}[!ht]
  \centering
  \caption{Recall@512 for different LoRA ranks. Best results are in
  \textbf{bold}, and second-best results are \underline{underlined}.}
  \label{tab:lora_rank}
  \small
  \setlength{\tabcolsep}{4pt}
  \renewcommand{\arraystretch}{1.10}
  \begin{tabular}{@{}c|cccc@{}}
    \toprule
    \textbf{Rank} & \textbf{Exp.} & \textbf{Long-View} & \textbf{Like} &
      \textbf{WT} \\
    \midrule
    4 & \underline{0.1562} & 0.1865 & \underline{0.1365} &
      \underline{0.2016} \\
    16 & 0.1543 & 0.1864 & 0.1355 & 0.1978 \\
    32 & \textbf{0.1565} & \textbf{0.1907} & \textbf{0.1391} &
      \textbf{0.2031} \\
    64 & 0.1559 & \underline{0.1892} & 0.1363 & 0.2010 \\
    \bottomrule
  \end{tabular}
\end{table}

\runinhead{Reward-History Size}
Table~\ref{tab:reward_group} varies the recent-history length used for
watch-time reward normalization. Increasing $K$ from 8 to 500 raises WT
Recall from 0.2003 to 0.2031 (+1.40\%) by stabilizing user-relative
watch-time preference estimates.

\begin{table}[!ht]
  \centering
  \caption{Effect of reward-history size on Watch-time Recall@512. Best result
  is in \textbf{bold}.}
  \label{tab:reward_group}
  \small
  \setlength{\tabcolsep}{4pt}
  \renewcommand{\arraystretch}{1.10}
  \begin{tabular}{@{}l|c@{}}
    \toprule
    \textbf{Reward-history size} & \textbf{WT Recall} \\
    \midrule
    $K=8$ & 0.2003 \\
    $K=500$ & \textbf{0.2031} \\
    \bottomrule
  \end{tabular}
\end{table}

\end{document}